# Addendum: Generalized Box-Müller Method for Generating $q$-Gaussian Random Deviates

Kenric P. Nelson[1], *Senior Member, IEEE* and William J. Thistleton[2]

***Abstract*— The generalized Box-Müller algorithm provides a methodology for generating $q$-Gaussian random variates. The parameter $-\infty < q \leq 3$ is related to the shape of the tail decay; $q < 1$ for compact-support including parabola $(q = 0)$; $1 < q \leq 3$ for heavy-tail including Cauchy $(q = 2)$. This addendum clarifies the transformation $q' = \frac{3q-1}{q+1}$ within the algorithm is due to a difference in the dimensions $d$ of the generalized logarithm and the generalized distribution. The transformation is clarified by the decomposition of $q = 1 + \frac{2\kappa}{1+d\kappa}$, where the shape parameter $-1 < \kappa \leq \infty$ quantifies the magnitude of the deformation from exponential. A simpler specification for the generalized Box- Müller algorithm is provided using the shape of the tail decay.**

***Index Terms*—Random Variables, Probability and Statistics, Information Theory, Complex Systems, Nonlinear, Pareto**

## I. Introduction

THISLETON, et. al. [1] described an algorithm for generating $q$-Gaussian random deviates by generalizing the Box-Müller method. The $q$-Gaussians generalize the Gaussian distribution $(q = 1)$ by modifying the tail decay to be either compact support $(q < 1)$, which is equivalent to the Pearson Type II distribution [2], or heavy-tail $(1 < q < 3)$, which is equivalent to the generalized Student's t, the Pearson Type VII, and Pareto Type IV [3] distributions. The algorithm specified a distinction between the parameter $q$ used to define the generalization of the logarithm function applied to two uniform random variates for the input and the parameter $q'$ which specified the $q$-Gaussian distribution of the output random variates. The algorithm consists of sampling two independent uniform random variables. The two uniform random variables are transformed by a generalized $q$-logarithm resulting in a two-dimensional $q$-Gaussian joint distribution. By extracting the one-dimensional marginal distribution, a $q'$-Gaussian random variate is generated. The random variate has a $q'$ value related to the $q$-logarithm by the relationship

$$q' = \frac{3q-1}{q+1}. \tag{1}$$

At the time of publication it was understood that $q$-statistic models often involved groups of parameters, another example being the $q$-triplet describing properties of the solar wind [4], but the precise reason for the different values of $q$ was not clear. An interpretation of $q$-statistics, which isolates the role of nonlinear coupling of statistical states, clarifies that the input $q$ parameter specifies a zero-dimensional logarithm, while the output $q'$ parameter is a one-dimensional distribution. An intermediate parameter, referred to here as $q''$, specifies a two-dimensional distribution. Since publication in 2007 [1], the generalized Box-Müller algorithm for generating $q$-Gaussians has been applied to a variety of applications [5]–[12] and further advances have been proposed [13].

First some background on the relationship between $q$-statistics and nonlinearity is provided. $q$-statistics has been used to model a variety of complex systems which exhibit nonextensive statistical mechanics [14]. A nonextensive system is one modeled by a generalized entropy that does not grow linearly for independent components, but instead includes a nonlinear term. The model is based on constraints using the $q^{\text{th}}$ power of the probability distribution $p^q(x)$. As such, $q$ quantifies the number of independent observations of the state $x$. Nevertheless, a variety of relationships can be established between $q$ and properties of complex systems distinct from linear systems including changes in the sensitivity to initial conditions, the distribution of stationary states, and relaxation associated with the autocorrelation function [15]. However, isolation of the role of nonlinearity in causing these changes requires decomposition of $q$ into its physical components.

Firstly, the $q$-logarithm and its inverse the $q$-exponential have definitions

$$\begin{aligned} \ln_q(x) &\equiv \frac{1}{1-q}\left(x^{1-q} - 1\right),\ x > 0 \\ \exp_q(x) &\equiv \left(1 + (1-q)x\right)_+^{\frac{1}{1-q}},\ (a)_+ \equiv \max(0, a) \end{aligned} \tag{2}$$

that emphasizes the significance of $r = 1 - q$, which will be referred to as the information risk bias [16] as (2) defines the risk aversion relative to the neutral measure of information $\ln(x)$. As will be discussed next this form of the generalization does not include the role of dimensions for a probability distribution and thus is equal to the zero-dimension case.

[1] Kenric P. Nelson is with Photrek, LLC, Watertown, MA 02472 (email: Kenric.Nelson@gmail.com).

[2] William J. Thistleton is with SUNY Polytechnic Institute, Utica, NY 13502 (email: thistlet@sunypoly.edu).

Multivariate analysis of the $q$-Gaussian distribution shows that the dimensionality $d$ and the power $\alpha$ of the state variable need to be isolated to specify the scale and shape of the distribution [17]. Following the definition of the multivariate t distribution without the normalization, but replacing the degree of freedom $\nu$ with its inverse $\kappa = 1/\nu$ which quantifies the nonlinearity gives a definition called the coupled Gaussian [17]

$$f(\mathbf{x}) \sim \left(1+\kappa\mathbf{x}^{\dagger}\Sigma^{-1}\mathbf{x}\right)_{+}^{-\frac{1}{2}\left(\frac{1}{\kappa}+d\right)}. \tag{3}$$

The domains of the coupled Gaussian are compact-support $-1/d \le \kappa < 0$, Gaussian $\kappa = 0$, and heavy-tail $\kappa > 0$. In contrast to (2) the distribution's exponent is not simply the inverse of the term multiplying the argument. The difference is due to the dimension $d$ of the distribution. In fact, although the cumulative distribution of the multivariate-$t$ is not analytic, its exponent has an asymptotic form of $x^{-\frac{1}{\kappa}}$. Thus the dimensional term is associated with the derivative of each dimension. The numeral 2 is associated with the power $\alpha$ of the variable for generalizations of the stretched-exponential distributions. Equating the exponents of (2) and (3) gives a decomposition of $q$ into three physical quantities

$$q = 1+\frac{\alpha\kappa}{1+d\kappa}., \tag{4}$$

The nonlinearity $\kappa$ is referred to as the *nonlinear statistical coupling* or simply the coupling. The coupling's inverse relationship with the degree of freedom is consistent with the need for more specifying parameters as the nonlinearity of the system increases. A number of other nonlinear physical properties have been equated to the coupling including the relative variance for superstatistics [18], the magnitude of multiplicative noise [19], and the complexity of social networks [20].

## II. Coupled Box-Müller Method

With the perspective of the nonlinear statistical coupling, the generalized Box-Müller method is now easier to express. The mean, scale, and shape of the desired coupled-Gaussian (or equivalently the $q$-Gaussian upon translation of the shape parameter) can be specified at the input and these parameters will remain invariant throughout the algorithm. The mean and scale have the same meaning as those for the Student's t distribution, but can also be defined in terms of their coupled moments which utilizes the coupled or escort probability (for the remainder of the text the subscript $\kappa$ on the moments will be dropped)

$$\mu \equiv \mu_{\kappa} \equiv \frac{\int\limits_{x\in X} x f^{1+\frac{2\kappa}{1+\kappa}}(x)\,dx}{\int\limits_{x\in X} f^{1+\frac{2\kappa}{1+\kappa}}(x)\,dx} \tag{5}$$

$$\sigma^2 \equiv \sigma_{\kappa}^2 \equiv \frac{\int\limits_{x\in X} (x-\mu)^2 f^{1+\frac{4\kappa}{1+\kappa}}(x)\,dx}{\int\limits_{x\in X} f^{1+\frac{4\kappa}{1+\kappa}}(x)\,dx} \tag{6}$$

$$\mu_n^n \equiv \mu_{\kappa,n}^n \equiv \frac{\int\limits_{x\in X} (x-\mu)^n f^{1+\frac{2n\kappa}{1+\kappa}}(x)\,dx}{\int\limits_{x\in X} f^{1+\frac{2n\kappa}{1+\kappa}}(x)\,dx} \tag{7}$$

Examining the generalized Box-Müller method in this context leads to the insight that the input q-logarithm is defined with $d = 0$, the intermediate 2-dimensional $q''$-Gaussian distribution is defined with $d'' = 2$ and the output is a one-dimensional $q'$-Gaussian distribution with $d' = 1$. In each case $\alpha = 2$ since the distributions of interest have a squared argument $x^2$ and the coupling $\kappa$ is invariant. The appendix provides a derivation of the $q$ relationships and main text defines the *coupled Box-Muller algorithm*.

The shape of the coupled Gaussian is defined by the coupling term. Examples include $\kappa = -1$ a uniform distribution over the scale $-\sigma < x < \sigma$; $\kappa = -1/3$ a compact-support with the shape of a parabola; $\kappa = 0$ the Gaussian with exponential decay; $\kappa = 1/2$ a heavy-tail at the boundary where the original variance diverges; $\kappa = 1$ is the Cauchy distribution and the boundary of divergent mean. The coupling range extends to infinity with increasingly slow decay of the coupled Gaussian tail.

From (3) we have the following definition for the multivariate coupled exponential function

$$\begin{aligned} &\exp_\kappa\left(x\right) \equiv \left(1+\kappa x\right)^{\frac{1}{\kappa}} \\ &\exp_\kappa^{-\frac{1+d\kappa}{2}}\left(\mathbf{x}^\dagger \Sigma^{-1}\mathbf{x}\right) = \left(1+\kappa \mathbf{x}^\dagger \Sigma^{-1}\mathbf{x}\right)^{-\frac{1}{2}\left(\frac{1+d\kappa}{\kappa}\right)}, \end{aligned} \tag{8}$$

and the zero-dimensional coupled logarithm is $\ln_\kappa x \equiv \frac{1}{\kappa}\left(x^\kappa - 1\right)$. Note that the $\frac{1}{2}$ in the exponent of (8) balances the $x^2$ terms in the argument so that the asymptotic behavior is proportional to $x^{-\left(\frac{1+d\kappa}{\kappa}\right)}$.

The coupled Box-Müller algorithm then has the following procedure.

1) Draw two uniform random variables $U_1$ and $U_2$ over the range 0 to 1.
2) Apply the generalized transformation

$$Z = \sqrt{\ln_\kappa U_1^{-2}}\cos\left(2\pi U_2\right). \tag{9}$$

A second complementary coupled Gaussian can also be generated, but unlike the Gaussian this second variable is not independent.

3) The location and scale for the coupled Gaussian variates are added to $Z$ by $X = \mu + \sigma Z$.

The observation that the bivariate pairs are uncorrelated but dependent relates to the properties of the joint distribution (3) of the coupled Gaussian. In this joint distribution there are no cross terms $z_1 z_2$ which would define a correlation; however, the distribution cannot be factored using the product function which is necessary for independence. Instead, the joint distribution factors via a generalization of the product function. Borges [21] defined the $q$-product based on the properties of the $q$-exponential

$$\begin{aligned} &e_q^x \otimes_q e_q^y \equiv e_q^{x+y} \\ &A \otimes_q B \equiv \left(A^{1-q} + B^{1-q}\right)^{\frac{1}{1-q}}; \end{aligned} \tag{10}$$

however, in order to model the product of marginal coupled exponential distributions forming an uncorrelated joint distribution the change in dimensionality needs to be accounted for. Thus a coupled-product defined based on the multi-dimensional properties of the coupled exponentials [17] needs to satisfy the following properties

$$\begin{aligned} &\exp_\kappa^{-\frac{1+\kappa}{\alpha}}\left(\left|x\right|^\alpha\right) \otimes_{-\alpha,\kappa} \exp_\kappa^{-\frac{1+\kappa}{\alpha}}\left(\left|y\right|^\alpha\right) \equiv \exp_\kappa^{-\frac{1+2\kappa}{\alpha}}\left(\left|x\right|^\alpha + \left|y\right|^\alpha\right) \\ &\text{where, } \exp_\kappa^\beta\left(x\right) \equiv \left(1+\kappa x\right)_+^{\frac{\beta}{\kappa}} \\ &A \otimes_{-\alpha,\kappa} B \equiv \left(A^{-\frac{\alpha\kappa}{1+\kappa}} + B^{-\frac{\alpha\kappa}{1+\kappa}}\right)^{-\frac{1+2\kappa}{\alpha\kappa}}. \end{aligned} \tag{11}$$

This dependency model is related to the $t$-Copula [22], though defined in terms of the probability distributions rather the cumulative distributions. The normalization terms of the distributions have to be treated separately.

Fig. *1* shows a standard Cauchy distribution with an empirical density formed from 10,000 random samples of the coupled Box-Müller method with $\mu = 0, \sigma = 1$ and $\kappa = 1$. Below the distributions is a scatter plot of the random variates. Included in the figure are arrows indicating the fluctuations that are proportional to the coupling. The coupled Gaussians can be derived from a Gaussian with a fluctuating scale $\theta$ such that $\theta^{-2}$ is distributed as a Gamma distribution. The coupling has thus been quantified to be the relative variance

$$\kappa = \frac{1}{2}\frac{E[\theta^{-4}] - E[\theta^{-2}]^2}{E[\theta^{-2}]^2} \tag{12}$$

Taking $\sqrt{\kappa}$ to be the relative standard deviation and multiplying it by the scale of the coupled Gaussian $\sigma$ gives a representation of the magnitude of the fluctuations. In [18] a sample estimator for scale was determined to be a function of the coupling and the geometric mean. Further investigation of the variance of the estimator could given insight into the magnitude of the sample fluctuations.

Fig. *2* shows theoretical and empirical distributions for a heavy-tail and compact-support coupled Gaussian with $\mu = 1$ and $\sigma = 2$. Below each distribution is a scatter plot of random variate samples. The heavy-tail distribution has coupling $\kappa = 0.5$ which is the boundary of infinity variance. A dual relationship exists between the compact-support and heavy-tail domain

$$\kappa \underset{CS}{\overset{HT}{\Leftrightarrow}} \frac{-\kappa}{1+d\kappa}. \tag{13}$$

An important set of these dual relationships for the coupled Gaussians are shown in Table 1. Thus the dual for the $\kappa = 0.5$ example is $\kappa = -\frac{1}{3}$. For the compact-support domain, the arrows representing the fluctuations, as shown in Fig. 2 (right), are in the opposite direction of the mean to scale indicating the reduction in entropy.

Table 1 Dual coupling values for compact-support and heavy-tail one-dimensional coupled Gaussians

| Compact-Support | Heavy-Tail |
|---|---|
| $-1/3$ | 0.5 |
| $-1/2$ | 1 |
| $-2/3$ | 2 |
| $-1$ | $\infty$ |

## III. Conclusion

The confirmation that the changes in the $q$ parameter for the generation of $q$-Gaussians using the generalized Box- Müller method is due to the dimensions of logarithm and exponential functions highlights the need to isolate the properties of complex systems which influence their statistical characteristics. While nonextensive statistical mechanics has contributed significant advances to the understanding of complex systems, a complete theory will require a decomposition of the $q$ parameter used in these models so that properties specific to the nonlinearity of complex dynamics can be isolated.

## IV. Appendix

First examine with transformation from the $q$-logarithm (2) input to the two-dimensional $q''$-Gaussian. The $q$-logarithm is used in the generalized Box-Mueller method to transform a pair of uniform random variables into a pair of $q$-Gaussians (see equation 16 and 17 of [1]). The resulting joint distribution for the two random variables $(Z_1, Z_2)$ is a bivariate Student's t

$$\begin{aligned} f_{Z_1,Z_2}(z_1,z_2) &= \frac{1}{2\pi} e_{(2-1/q)}^{-\frac{q}{2}\left(z_1^2+z_2^2\right)} \\ e_q^x &\equiv \left(1+(1-q)x\right)_+^{\frac{1}{1-q}} \\ (a)_+ &\equiv \max(0,a). \end{aligned} \tag{14}$$

The parameter $q'' = 2 - \frac{1}{q}$ can be shown to be a transformation from a zero-dimensional specification for the logarithm to a two-dimensional specification for the joint distribution. Solving for $d''$ utilizing (4) gives

$$\begin{aligned} q'' &= 2 - \tfrac{1}{q} \\ 1 + \frac{2\kappa}{1+d''\kappa} &= 2 - \left(1 + \frac{2\kappa}{1+(d=0)\kappa}\right)^{-1} \\ \frac{2\kappa}{1+d''\kappa} &= 1 - \frac{1}{1+2\kappa} \\ 1 + d''\kappa &= 2\kappa\left(\frac{1+2\kappa}{2\kappa}\right) \\ d'' &= 2 \end{aligned} \tag{15}$$

The individual variates are formed via integration of the two-dimensional distribution into one-dimensional marginal distributions. This can be shown to be a specification that $d'=1$ from the relationship observed between the input and output parameters $q'=\frac{3q-1}{q+1}$

$$
\begin{aligned}
1+\frac{2\kappa}{1+d'\kappa} &= \frac{3\left(1+\frac{2\kappa}{1+(d=0)\kappa}\right)-1}{1+\frac{2\kappa}{1+(d=0)\kappa}+1} \\
\frac{1}{1+d'\kappa} &= \frac{1}{2\kappa}\left(\frac{2+6\kappa}{2+2\kappa}-1\right) \\
1+d'\kappa &= 2\kappa\left(\frac{2+2\kappa}{4\kappa}\right) \\
d' &= 1
\end{aligned} \tag{16}
$$

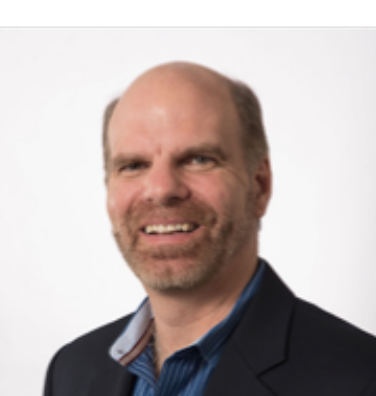

**Kenric P. Nelson** (M’85–SM’09) received degrees in electrical engineering including a B.S. degree *summa cum laude* from Tulane University, a M.S. degree from Rensselaer Polytechnic Institute, and a Ph.D. degree from Boston University. His education in Program Management includes an Executive Certificate from MIT Sloan and certification with the Program Management Institute.

He is President and Founder of Photrek, LLC which provides scientific and engineering services for the design and analysis of complex systems. Previously, he was a Research Professor with Boston University Electrical & Computer Engineering and a Sr. Principal Systems Engineer with Raytheon Company.

He has pioneered novel approaches to measuring and fusing information in complex decision systems. His *nonlinear statistical coupling* methods have been used to improve the accuracy and robustness of radar signal processing, sensor fusion, and machine learning algorithms.

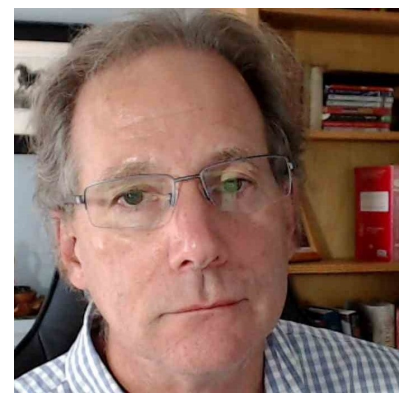

**William Thistleton** received his BS in Electrical Engineering from Clarkson University, an MA in Mathematics from SUNY Potsdam and a PhD in Applied Mathematics (1996) from Stony Brook University.

He is Associate Professor of Mathematics at SUNY Polytechnic Institute, which he joined in1993. His research and consulting interests in applied statistics and data analysis including price modeling, the effects of architecture on patient care in long term care facilities, psychosocial aspects of Fibromyalgia, and the development of image processing algorithms on Quantum Annealing platforms. He has developed and implemented algorithms for use in bitstream analysis, numerical methods, and analysis of correlated systems with an emphasis on non-extensive methods.

He is a passionate and innovative educator, having authored corporate training courses in data analysis and a popular MOOC in Time Series Analysis, in addition to traditional and distance classes in mathematics and statistics.

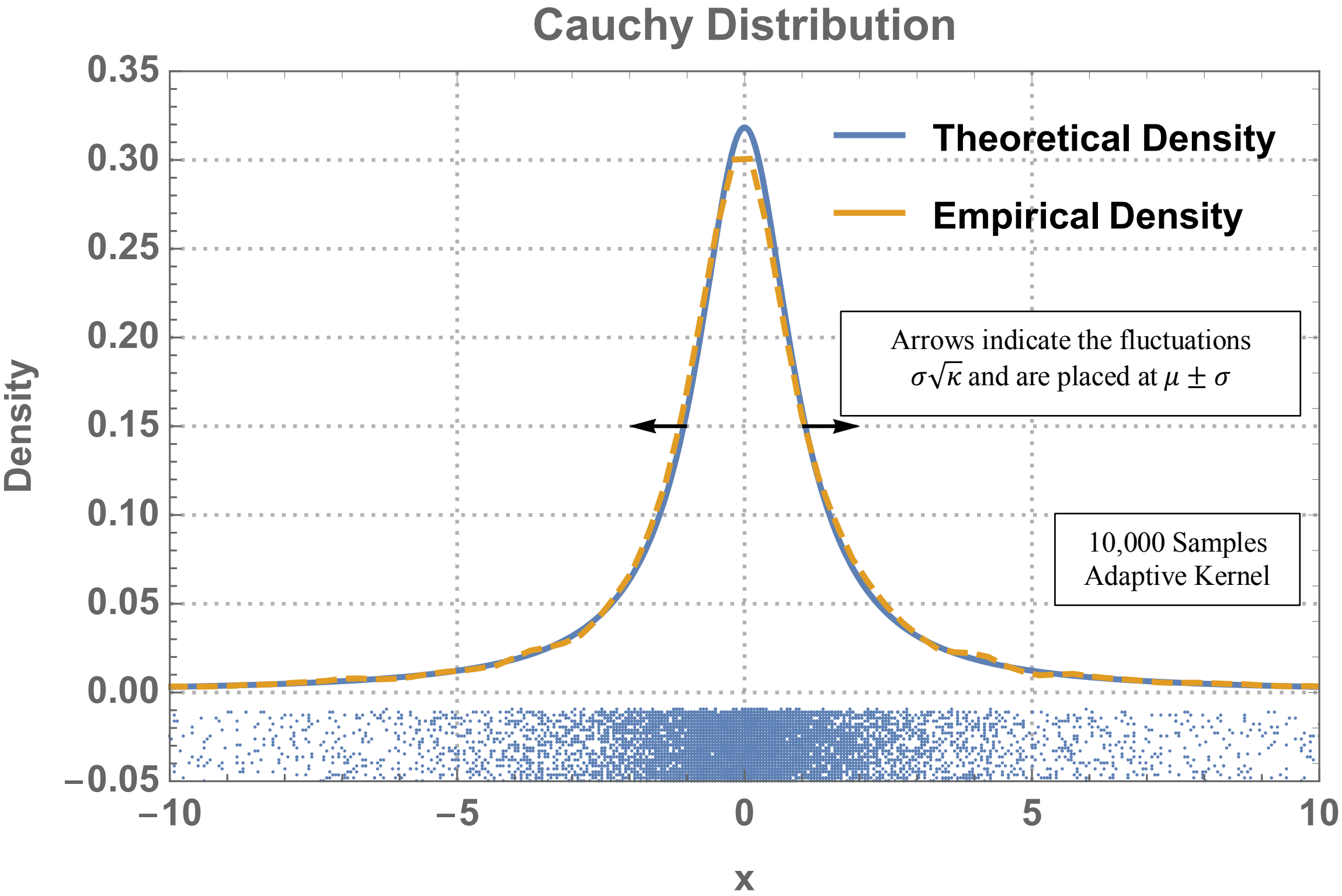

Fig. 1. The Cauchy distribution showing an theoretical and empirical probability density function. The empirical distribution is formed using the Silverman method. Below the distribution is a scatter plot of the 10,000 samples (excluding those outside the plots domain). The Cauchy distribution, shown with zero mean and a scale of 1, has a coupling value of 1. The arrows of length 1 represent the magnitude of the fluctuation of the scale and are proportional to $\sigma\sqrt{\kappa}$ .

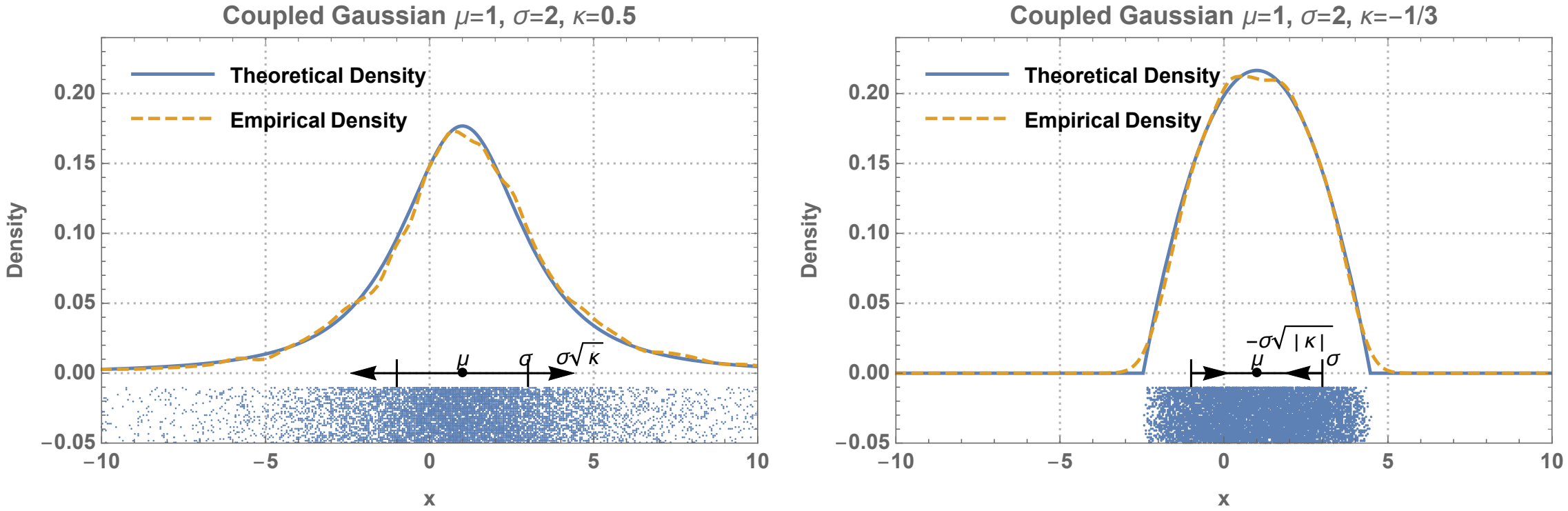

Fig. 2. Theoretical and empirical probability densities for coupled Gaussian distributions in the a) heavy-tail and b) compact-support domain. Below the distributions are scatter plots of the random variates. Both distributions have $\mu = 1$ and $\sigma = 2$. The heavy-tail distribution has coupling of $\kappa = 0.5$ which is the boundary of infinite standard deviation. The compact-support is the conjugate value of $\kappa = -{}^{1}/_{3}$. Below the densities the mean, scale and fluctuations are represented. For the compact-support the fluctuations reduce the entropy of the distribution. The empirical distribution is based on an adaptive bandwidth kernel of 10,000 samples.

# Generalized Box-Müller Method for Generating *q*-Gaussian Random Deviates

William J. Thistleton, John A. Marsh, *Member IEEE*
Kenric Nelson, *Member IEEE* and Constantino Tsallis

*Abstract*— **The *q*-Gaussian distribution is known to be an attractor of certain correlated systems and is the distribution which, under appropriate constraints, maximizes a generalization of the familiar Shannon entropy. This generalized entropy, or *q*-entropy, provides the basis of non-extensive statistical mechanics, a theory which is postulated as a natural extension of the standard (Boltzmann-Gibbs) statistical mechanics, and which may explain the ubiquitous appearance of heavy-tailed distributions in both natural and man-made systems. The *q*-Gaussian distribution is also used as a numerical tool, for example as a visiting distribution in Generalized Simulated Annealing. A simple, easy to implement numerical method for generating random deviates from a *q*-Gaussian distribution based upon a generalization of the well known Box-Müller method is developed and presented. This method is suitable for a larger range of *q* values, $-\infty < q < 3$, than has previously appeared in the literature, and can generate deviates from *q*-Gaussian distributions of arbitrary width and center. MATLAB code showing a straightforward implementation is also included.**

*Index Terms*—**Entropy Maximizing Distribution, q-Gaussian Distribution, Random Number Generation.**

## V. Introduction

The Gaussian (normal) distribution is ubiquitous in probability and statistics due to its role as an attractor of independent systems with finite variance. It is also the distribution which maximizes the Boltzmann-Gibbs entropy $S_{BG} \equiv E\left[-\log\left(f\left(X\right)\right)\right] = -\int_{-\infty}^{\infty}\log\left[f(x)\right]f(x)dx$ under appropriate constraints [i]. There are many methods available to transform computed pseudorandom uniform deviates into normal deviates, for example the "ziggurat method" of Marsaglia and Tsang [ii]. One of the most popular methods is that of Box and Müller [iii]. Though not optimal in its properties, it is easy to understand and to implement numerically.

Since *q*-Gaussian distributions are ubiquitous within the framework of nonextensive statistical mechanics [iv] and are applicable to a variety of complex signals and systems, there is a need for a *q*-Gaussian generator. The *q*-Gaussian distribution was originally defined as the maximizing distribution for the entropic form

$$S_q = \frac{1-\int_{-\infty}^{\infty}\left[p\left(x\right)\right]^q dx}{q-1} \quad \left(q \in \Re\right) \qquad (1)$$

under appropriate constraints [v]. Generalizations of Gauss' Law of Errors [vi] and the Central Limit Theorem [vii] further demonstrate the underlying prevalence for complex systems to exhibit *q*-Gaussian characteristics. Mechanisms which have been shown to produce *q*-Gaussian distributions are reviewed in [viii] and include multiplicative noise, weakly chaotic dynamics, correlated anomalous diffusion, preferential growth of networks, and asymptotically scale-invariant correlations. The *q*-Gaussian distributions have been applied in a broad range of fields [ix] including thermodynamics, biology, economics, and quantum mechanics. Within electrical and computing systems, applications have included wavelet signal processing [x], generalized simulated annealing [xi], quantum information theory [xii], and image fusion [xiii]. Other areas worth further investigation are communications systems and coding techniques which are influenced by multiplicative noise or exhibit power-law distributions.

A recent paper by Deng et. al. [xiv] specifies an algorithm to generate *q*-Gaussian random deviates as the ratio of a standard normal distribution and square root of an independent $\chi^2$ scaled by its degrees of freedom. However, since the resulting Gosset's Student-*t* distribution is a special case of the *q*-Gaussian for *q* in the range 1 to 3, our method allows the generation of a fuller set of distributions [xv]. Bailey's Method [xvi] for generating the Student-*t* agrees with our method for positive values of the degrees of freedom parameter but does not produce values for compact support *q*-Gaussian distributions.

Manuscript received June 8, 2006 The authors gratefully acknowledge support from A. Williams of the Air Force Research Laboratory, Information Directorate, and SI International, under contract No. FA8756-04-C-0258.

W. J. Thistleton is with Department of Mathematics, SUNY Institute of Technology, Utica NY, 13504, USA, (email:William.Thistleton@sunyit.edu).

J. A. Marsh, is with Assured Information Security, Inc, 245 Hill Road, Rome, NY 13442, USA (e-mail: John.Marsh@ainfosec.com).

K. Nelson is a Consultant, 100 Richardson Rd., Hollis, NH 03049, USA (e-mail: Kenric.Nelson@ieee.org).

C. Tsallis is with Santa Fe Institute, 1399 Hyde Park Road, Santa Fe, NM 87501, USA and Centro Brasileiro de Pesquisas Fisicas Xavier Sigaud 150, 22290-180 Rio de Janeiro-RJ, Brazil (e-mail: tsallis@santafe.edu).

In this paper, we generalize the Box-Müller algorithm to produce $q$-Gaussian deviates. Our technique works over the full range of interesting $q$ values, from -∞ to 3. In what follows, we show analytically that the algorithm produces random numbers drawn from the correct $q$-Gaussian distribution and we demonstrate the algorithm numerically for several interesting values of $q$. We note here the fact that, while the original Box-Müller algorithm produces pairs of independent Gaussian deviates, our technique produces $q$-Gaussian deviate pairs that are uncorrelated but not independent. Indeed, for $q<1$ the joint probability density function of the $q$-Gaussian has support on a circle while the marginal distributions have support on intervals. Thus the joint distribution may not be recovered simply as the product of the marginal distributions.

## VI. Review of the Box Müller Technique

We first review the well known method of Box-Müller for generating Gaussian deviates from a standard normal distribution. We then present a simple variation of this method useful for generating $q$-Gaussian deviates. Our method is easily understood and can be coded very simply using the pseudo-random number generator for uniform deviates available in almost all computational environments.

Box and Müller built their method from the following observations. Given $Z_1, Z_2 \overset{iid}{\approx} N(0,1)$ uncorrelated (hence in this case independent) standard normal distributions with mean 0 and variance 1, contours of their joint probability density function are circles with $\Theta \equiv \arctan(Z_1/Z_2)$ uniformly distributed. Also, the square of a standard normal distribution has a $\chi^2$ distribution with 1 degree of freedom. Denote a $\chi^2$ random variable with ν degrees of freedom as $\chi^2_\nu$ and write $Z^2 \sim \chi^2_1$. Because the sum of two independent $\chi^2$ distributions each with 1 degree of freedom is again $\chi^2$ but with 2 degrees of freedom, the random variable defined as $R^2 = Z_1^2 + Z_2^2$ has a $\chi^2_2$ distribution. Related to this point, if a random variable $R^2$ has a $\chi^2_2$ distribution then $U \equiv \exp(-\frac{1}{2}R^2)$ is uniformly distributed. We are led to the Box-Müller transformations

$$
\begin{aligned}
U_1 &= \exp\left(-\frac{1}{2}\left(Z_1^2 + Z_2^2\right)\right) \\
U_2 &= \frac{1}{2\pi}\operatorname{atan}\left(\frac{Z_2}{Z_1}\right)
\end{aligned}
\qquad (2a,b)
$$

and inverse transformations

$$Z_1 = \sqrt{-2\ln\left(U_1\right)}\cos\left(2\pi U_2\right), \quad (3)$$

and

$$Z_2 = \sqrt{-2\ln\left(U_1\right)}\sin\left(2\pi U_2\right). \quad (4)$$

We now apply these transformations assuming $U_1$ and $U_2$ are independent, uniformly distributed on the interval $(0,1)$, and show that $Z_1$ and $Z_2$ are normally distributed. Indeed, given a transformation $\left(Z_1, Z_2\right) = T\left(U_1, U_2\right)$ between pairs of random variables, we construct the joint density of $\left(Z_1, Z_2\right)$, denoted $f_{Z_1,Z_2(z_1,z_2)}$, as

$$f_{Z_1,Z_2}\left(z_1,z_2\right) = f_{U_1,U_2}\left(z_1,z_2\right)\left|J\left(z_1,z_2\right)\right| \quad (5)$$

where the Jacobian is given by

$$J\left(z_1,z_2\right) \equiv \frac{\partial u_1}{\partial z_1}\frac{\partial u_2}{\partial z_2} - \frac{\partial u_1}{\partial z_2}\frac{\partial u_2}{\partial z_1}. \quad (6)$$

Since $U_1$ and $U_2$ are independent, uniformly distributed random variables on $(0,1)$ we immediately have $f_{U_1,U_2}\left(z_1,z_2\right) = 1$. Evaluation of the Jacobian yields

$$
\begin{aligned}
f_{Z_1,Z_2}\left(z_1,z_2\right) &= \frac{1}{2\pi}\exp\left(-\left(z_1^2 + z_2^2\right)/2\right) \\
&= \left[\frac{1}{\sqrt{2\pi}}\exp\left(-z_1^2/2\right)\right]\left[\frac{1}{\sqrt{2\pi}}\exp\left(-z_2^2/2\right)\right].
\end{aligned}
\quad (7)
$$

The Box-Müller method is thus very simple: start with two independent random numbers drawn from the range $(0,1)$, apply the transformations of (3)-(4), and one arrives at two random numbers drawn from a standard Gaussian distribution (mean 0 and variance 1). The two random deviates thus generated are independent due to the separation of the terms in (7).

## VII. The $Q$-Gaussian Distribution

Before presenting our generalization of the Box-Müller algorithm, we briefly review the definition and properties of the $q$-Gaussian function. First, we introduce the $q$-logarithm and its inverse, the $q$-exponential [xvii], as

$$\ln_q(x) \equiv \frac{x^{1-q} - 1}{1-q} \qquad x > 0 \quad (8)$$

and

$$e_q^x \equiv \begin{cases} \left[1+(1-q)x\right]^{\frac{1}{(1-q)}} & 1+(1-q)x \geq 0 \\ 0 & else \end{cases}. \quad (9)$$

These functions reduce to the usual logarithm and exponential functions when $q=1$. The $q$-Gaussian density is defined for $-\infty < q < 3$ as

$$
\begin{aligned}
p\left(x;\bar{\mu}_q,\bar{\sigma}_q\right) &= A_q\sqrt{B_q}\left[1+\left(q-1\right)B_q\left(x-\bar{\mu}_q\right)^2\right]^{1/(1-q)} \\
&= A_q\sqrt{B_q}\,e_q^{-B_q\left(x-\bar{\mu}_q\right)^2}
\end{aligned}
. \quad (10)
$$

where the parameters $\bar{\mu}_q$, $\bar{\sigma}_q^2$, $A_q$, and $B_q$ are defined as follows. First, the $q$-mean $\bar{\mu}_q$ is defined analogously to the usual mean, except using the so-called $q$-expectation value (based on the escort distribution), as follows

$$\bar{\mu}_q \equiv \langle x \rangle_q \equiv \frac{\int x \left[ p(x) \right]^q dx}{\int \left[ p(x) \right]^q dx}. \qquad (11)$$

Similarly, the $q$-variance, $\bar{\sigma}_q^2$ is defined analogously to the usual second order central moment, as

$$\bar{\sigma}_q^2 \equiv \left\langle \left( x - \bar{\mu}_q \right)^2 \right\rangle_q \equiv \frac{\int \left( x - \bar{\mu}_q \right)^2 \left[ p(x) \right]^q dx}{\int \left[ p(x) \right]^q dx}. \qquad (12)$$

When $q = 1$, these expressions reduce to the usual mean and variance. The normalization factor is given by

$$A_q = \begin{cases} \dfrac{\Gamma\left[\dfrac{5-3q}{2(1-q)}\right]}{\Gamma\left[\dfrac{2-q}{1-q}\right]} \sqrt{\dfrac{1-q}{\pi}} & q < 1 \\ \dfrac{1}{\sqrt{\pi}} & q = 1 \\ \dfrac{\Gamma\left[\dfrac{1}{q-1}\right]}{\Gamma\left[\dfrac{3-q}{2(q-1)}\right]} \sqrt{\dfrac{q-1}{\pi}} & 1 < q < 3 \end{cases}. \qquad (13)$$

Finally, the width of the distribution is characterized by

$$B_q = \left[ (3-q) \bar{\sigma}_q^2 \right]^{-1} \qquad q \in (-\infty, 3). \qquad (14)$$

Denote a general $q$-Gaussian random variable $X$ with $q$-mean $\bar{\mu}_q$ and $q$-variance $\bar{\sigma}_q$ as $X \sim N_q\left(\bar{\mu}_q, \bar{\sigma}_q^2\right)$, and call the special case of $\bar{\mu}_q \equiv 0$ and $\bar{\sigma}_q^2 \equiv 1$ a *standard q-Gaussian,* $Z \sim N_q(0,1)$. The density of the standard $q$-Gaussian distribution may then be written as

$$p\left(x; \bar{\mu}_q = 0, \bar{\sigma}_q = 1\right) = A_q \sqrt{B_q}\, e_q^{-B_q x^2} = \frac{A_q}{\sqrt{3-q}} \left[ 1 + \frac{q-1}{3-q} x^2 \right]^{\frac{1}{1-q}} \qquad (15)$$

We show below (23) that the marginal distributions associated with our transformed random variables recover this form with new parameter $q'$.

The $q$-Gaussian distribution reproduces the usual Gaussian distribution when $q = 1$. Its density has compact support for $q < 1$, and decays asymptotically as a power law for $1 < q < 3$. For $3 \le q$, the form given in (10) is not normalizable. The usual variance (second order moment) is finite for $q < 5/3$, and, for the standard $q$-Gaussian, is given by $\sigma^2 = (3-q)/(5-3q)$. The usual variance of the $q$-Gaussian diverges for $5/3 \le q < 3$, however the $q$-variance remains finite for the full range $-\infty < q < 3$, equal to unity for the standard $q$-Gaussian.

## VIII. Generalized Box MÜller for Q-Gaussian Deviates

Our generalization of the Box-Müller technique is based on preserving its circular symmetry, while changing the behavior in the radial direction. Our algorithm starts with the same two uniform deviates as the original Box-Müller technique, and applies a similar transformation to yield $q$-Gaussian deviates. Thus it preserves the simplicity of implementation that makes the original Box-Müller technique so useful.

Indeed, given independent uniform distributions $U_1$ and $U_2$ defined on $(0,1)$, we define two new random variables $Z_1$ and $Z_2$ as

$$Z_1 \equiv \sqrt{-2 \ln_q \left( U_1 \right)} \cos\left( 2\pi U_2 \right) \qquad (16)$$

and

$$Z_2 \equiv \sqrt{-2 \ln_q \left( U_1 \right)} \sin\left( 2\pi U_2 \right). \qquad (17)$$

We now show that each of $Z_1$ and $Z_2$ is a standard $q$-Gaussian characterized by a new parameter $q' = \frac{3q-1}{q+1}$. Thus as $q$ (used in the $q$-log of the transformation) ranges over $(-1, \infty)$, the $q$-Gaussian is characterized by $q'$ in the range $(-\infty, 3)$, which is the interesting range for the $q$-Gaussian. For $q' \ge 3$ the distribution can not be normalized.

Proceed as follows. Obtain the inverse transformations as

$$U_1 = \exp_q\left( -\frac{1}{2}\left( Z_1^2 + Z_2^2 \right) \right) \qquad (18)$$

and

$$U_2 = \frac{1}{2\pi} \operatorname{atan}\left( \frac{Z_2}{Z_1} \right), \qquad (19)$$

so that the Jacobian is

$$J\left( z_1, z_2 \right) = -\frac{1}{2\pi} \left[ \exp_q\left( -\frac{1}{2}\left( z_1^2 + z_2^2 \right) \right) \right]^q. \qquad (20)$$

And, since $f_{U_1,U_2}(u_1,u_2) = 1$, we obtain the joint density of $Z_1$ and $Z_2$ as

$$f_{Z_1,Z_2}\left( z_1, z_2 \right) = \frac{1}{2\pi} \left( e_q^{-\frac{1}{2}\left( z_1^2 + z_2^2 \right)} \right)^q = \frac{1}{2\pi} e_{(2-1/q)}^{-\frac{q}{2}\left( z_1^2 + z_2^2 \right)}. \qquad (21)$$

Note that for the second part of the above equality we have manipulated the definition of the $q$-exponential as $\left[ e_q^x \right]^q = \left[ 1 + (1-q)x \right]^{\frac{q}{1-q}} = \left[ 1 + (1-p)qx \right]^{\frac{1}{1-p}}$ for $p = 2 - \frac{1}{q}$. Note also that in the limit as $q \to 1$ we recover the product of independent standard normal distributions as required.

The marginal distributions resulting from this joint density are standard $q$-Gaussian. As illustrated below for the case $q > 1$, one obtains the marginal distributions by integrating (with the aid of the Maple symbolic integration facility) the joint density

$$
\begin{aligned}
I &= \frac{1}{2\pi}\int_{-\infty}^{\infty} f_{Z_1,Z_2}\left(z_1,z_2\right)\,dz_1 \\
&= \frac{1}{\sqrt{\pi}}\frac{\Gamma\left(\dfrac{q+1}{2(q-1)}\right)}{\Gamma\left(\dfrac{1}{q-1}\right)}\sqrt{\frac{1}{2}(q-1)}\left(1-\frac{1}{2}(1-q)\,z_2^2\right)^{\frac{1}{2}\left(\frac{1+q}{1-q}\right)}
\end{aligned}
\qquad (22)
$$

Note that, by the definition of $q'$, we have $\frac{1}{2}(q-1)=\frac{q'-1}{3-q'}$ and $\frac{1}{2}\left(\frac{1+q}{1-q}\right)=\frac{1}{1-q'}$, and so we obtain

$$
\begin{aligned}
I &= \frac{1}{\sqrt{\pi}}\frac{\Gamma\left(\dfrac{1}{q'-1}\right)}{\Gamma\left(\dfrac{3-q'}{2(q'-1)}\right)}\sqrt{\frac{q'-1}{3-q'}}\left(1+\frac{q'-1}{3-q'}x^2\right)^{\frac{1}{1-q'}} \\
&= A_{q'}\sqrt{B_{q'}}\,e_{q'}^{-B_{q'}x^2}
\end{aligned}
\qquad (23)
$$

As seen from (15) this is a standard $q$-Gaussian with parameter $q'$.

An important result is that the standard $q$-Gaussian recovers the Student's-t distribution with $\eta$ degrees of freedom when $q\equiv(3+\eta)/(1+\eta)$. Therefore, our method may also be used to generate random deviates from a Student's-*t* distribution by taking $q'\equiv\frac{3+\eta}{1+\eta}$, or $q=\frac{q'+1}{3-q'}=\frac{2+\eta}{\eta}$. Also, it is interesting to note that, in the limit as $q'\to-\infty$ the $q$-Gaussian recovers a uniform distribution on the interval $(-1,1)$.

To illustrate our method, we consider the following empirical probability density functions (relative frequency histograms normalized to have area equal to 1) of generated deviates, with $q$-Gaussian density functions superimposed. Each histogram is the result of $5\times10^6$ deviates. We present first two typical histograms for $q'<1$. Fig. 1 shows the case $q'=-5.0$ and Fig. 2 shows the case $q'=1/7$. Both the empirical densities show excellent agreement with the theoretical $q$-Gaussian densities, which have compact support since $q'<1$. For $q'>1$, the density has heavy tails and so a standard empirical probability density function is of little value. We show three different plots of the empirical density for the case $q'=2.75$. In Fig. 3 we show a semi-log plot of the empirical density function, which helps to illustrate the behavior in the near tails of the distribution. In Fig. 4 we plot the $q$-log of the (appropriately scaled) density versus $x^2$, so the resulting density function appears as a straight line with slope equal to $-B_q$. Fig. 5 provides another validation of the method. We transform as $Y=\log(|X|)$ to produce a random variable without heavy tails, which may be treated with a traditional histogram. This random variable has density

$$
f_Y\left(y\right)=2e^{y}A_q\sqrt{B_q}\left[1-\frac{1}{2}(1-q)e^{2y}\right]^{\frac{1}{2}\cdot\left(\frac{1+q}{1-q}\right)}. \qquad (24)
$$

We demonstrate the utility of the $q$-variance $\overline{\sigma}_q^2$ as a characterization of heavy tail $q$-Gaussian distributions in Fig. 6. For standard $q$-Gaussian data generated for various values of parameter $q$, we see that as $q'\to5/3$ the second order central moment (i.e., the usual variance) diverges, whereas the $q$-variance remains steady at unity. For $q'\approx3$ numerical determination of $\overline{\sigma}_q^2$ is difficult.

The algorithm described in equations (16)-(17) yields random deviates drawn from a standard $q$-Gaussian distribution. To produce deviates drawn from a general $q$-Gaussian distribution with $q$-mean $\overline{\mu}_q$ and $q$-variance $\overline{\sigma}_q^2$, one simply performs the usual transformation $X\equiv\overline{\mu}_q+\overline{\sigma}_q Z$. Here $Z\sim N_q(0,1)$ results in $X\sim N_q\left(\overline{\mu}_q,\overline{\sigma}_q^2\right)$. Fig. 7 illustrates this transformation, and also further illustrates the utility of the $q$-variance as a method of characterization. Shown are the variance and $q$-variance of sample data obtained by transforming standard $q$-Gaussian data, $Z\sim N_q(0,1)$ with $q'=1.4$, via $X=5+3Z\sim N_q(5,9)$. Data were generated for sample sizes $N_S$ ranging from 50 to 2000. As the sample size grows, the $q$-variance converges nicely to the expected value $\overline{\sigma}_q^2=9$. The sample variances, however, converge relatively slowly to the predicted value $\sigma^2=\left[(3-q')/(5-3q')\right]\sigma_q^2=18$ (where the quantity $(3-q')/(5-3q')=2$ is the variance of the untransformed data). The utility of the $q$-variance is illustrated even more dramatically in Fig. 8 where $q'=2.75$. Note the variance diverges as $N_S$ grows, whereas the $q$-variance remains at the predicted value of $\overline{\sigma}_q^2=9$.

## IX. SUMMARY AND CONCLUSION

We have presented a method for generating random deviates drawn from a $q$-Gaussian distribution with arbitrary spread and center, and over the full range of meaningful $q$ values $-\infty<q<3$. The method presented is simple to code and relatively efficient. The relationship between generating $q$ value and target $q'$ value was developed and verified.

It should be noted that the traditional Box-Müller method produces uncorrelated pairs of bivariate normal deviates. For the case of bivariate normality this implies that the deviates so produced are generated in independent pairs. For the $q$-Gaussian, however, even when the second order moment exists (i.e., $q$ < 5/3) the deviate pairs are uncorrelated but not independent. It is an interesting question, unresolved at this point, as to whether these pairs are $q$-correlated [vii]. Random variables which are $q$-correlated have a special correlation structure in that the $q$-Fourier Transform of the sum of the random variables must be the *q*-product of the individual *q*-Fourier Transforms.

Applications of this *q*-Gaussian random number generator include many numerical techniques for which a heavy-tailed (q > 1) or compact-support (q < 1) distribution is required. Examples include generating the visiting step sizes in generalized simulated annealing and generating noise sources with long-range correlations.

## APPENDIX

We present, for convenience of implementation, a MATLAB code for the generalized Box-Müller method presented herein. The code shown is intended to demonstrate the algorithm, and is not optimized for speed.

The algorithm is straightforward to implement, and is shown below as two functions. The first function `qGaussianDist` generates the *q*-Gaussian random deviates, and calls the second function, `log_q`, which calculates the *q*-log. The method relies on a high quality uniform random number generator that produces deviates in the range (0,1). The MATLAB command producing these deviates is called `rand`, and in the default implementation in version 7.1, uses a method due to Marsagalia [xviii]. The other MATLAB-specific component to the code given below is the built-in value `eps`, whose value is approximately `1E-16` and which represents the limit of double precision.

```
function qGaussian  = qGaussianDist(nSamples,qDist)
%
% Returns random deviates drawn from a q-gaussian
% distribution.
% The number of samples returned is nSamples.
% The q that characterizes the q-Gaussian is given
% by qDist
%
    % Check that q < 3
    if qDist < 3
        % Calculate the q to be used on the q-log
        qGen = (1 + qDist)/(3 - qDist);

        % Initialize the output vector
        qGaussian = zeros(1,nSamples);

        % Loop through, populate the output vector
        for k = 1:nSamples
            % Get two uniform random deviates
            % from built-in rand function
            u1 = rand;
            u2 = rand;

            % Apply the q-Box-Muller algorithm,
            % taking only one of two possible values
            R= sqrt(-2*log_q(u1,qGen))
            qGaussian(k)=R*sin(2*pi*u2);
        end

    % Return 0 and give a warning if q >= 3
    else
        warning('q value must be less than 3')
        qGaussian = 0;
    end

end
```

```
function a = log_q(x,q)
%
% Returns the q-log of x, using q
%

    % Check to see if q = 1 (to double precision)
    if abs(q - 1) < 10*eps

      % If q is 1, use the usual natural logarithm
      a = log(x);

    else

      % If q differs from 1, use def of the q-log
      a = (x.^(1-q) - 1)./(1-q);

    end

end
```

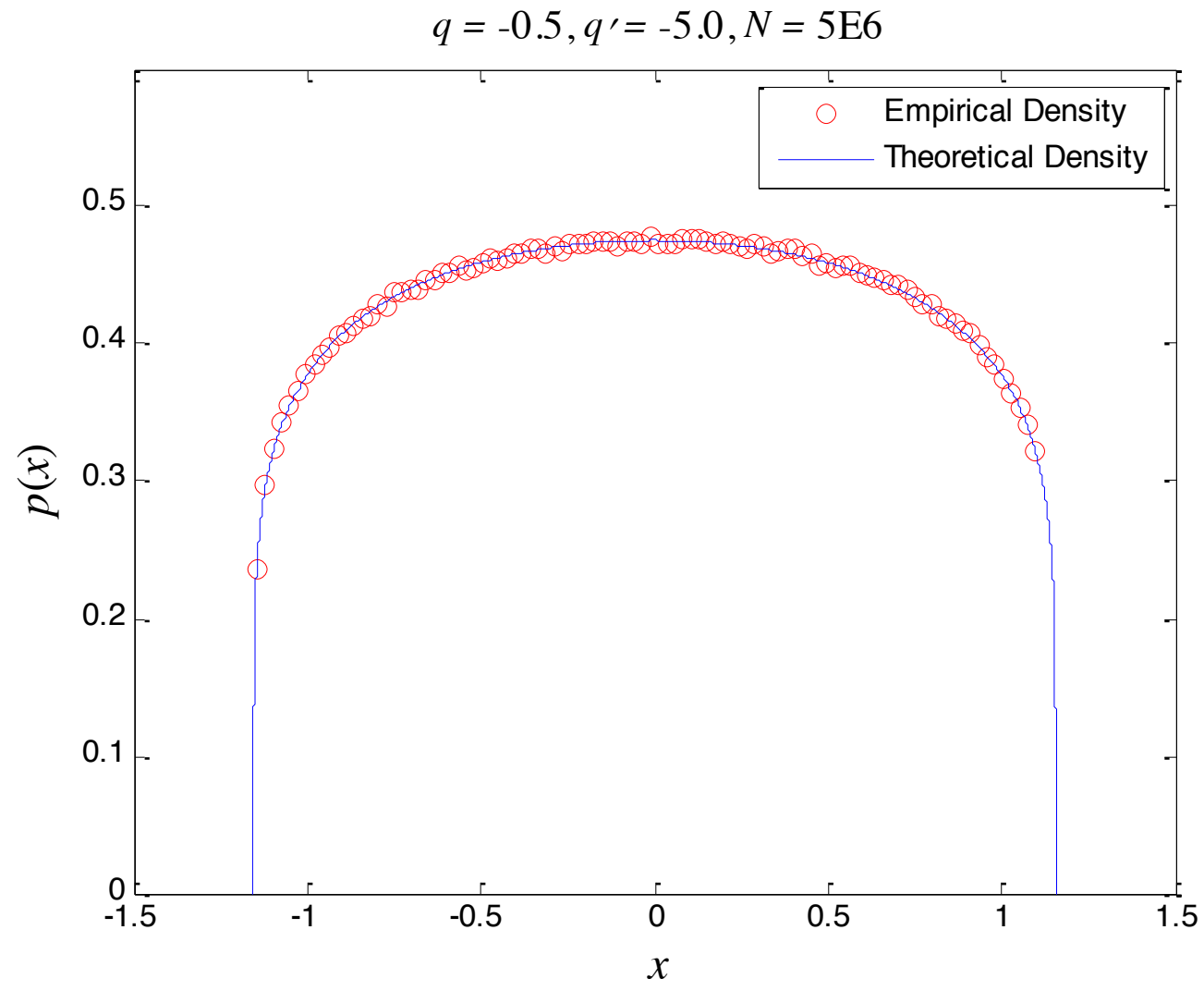

*Figure 1: Theoretical density* p(x) *and histogram of simulated data for a standard* q*-Gaussian distribution with* $q' < 0$. *The distribution has compact support for this* $q'$.

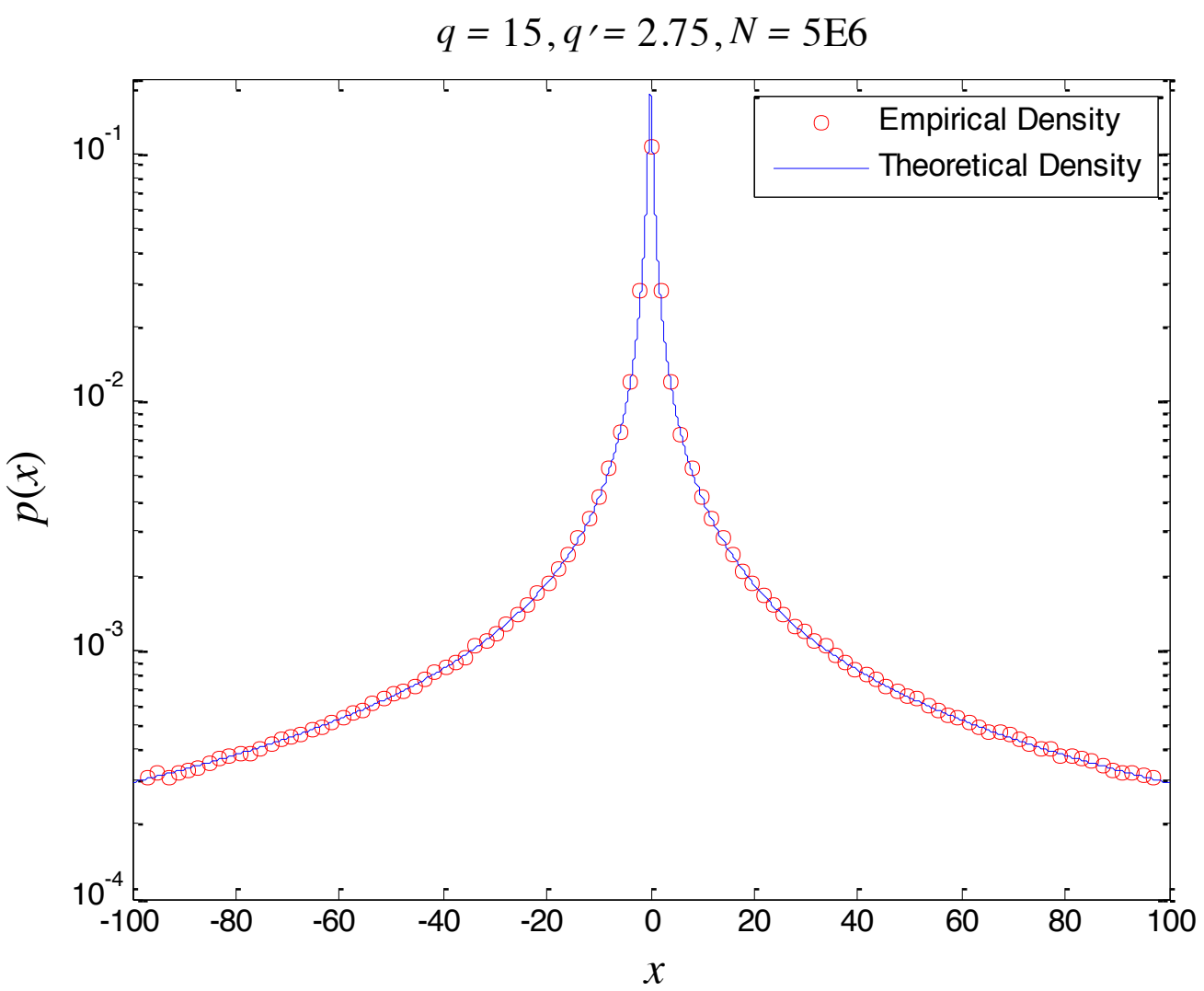

*Figure 3: Theoretical density* p(x) *and histogram of simulated data for a standard* q*-Gaussian distribution with* $1 < q' < 3$, *truncated domain.*

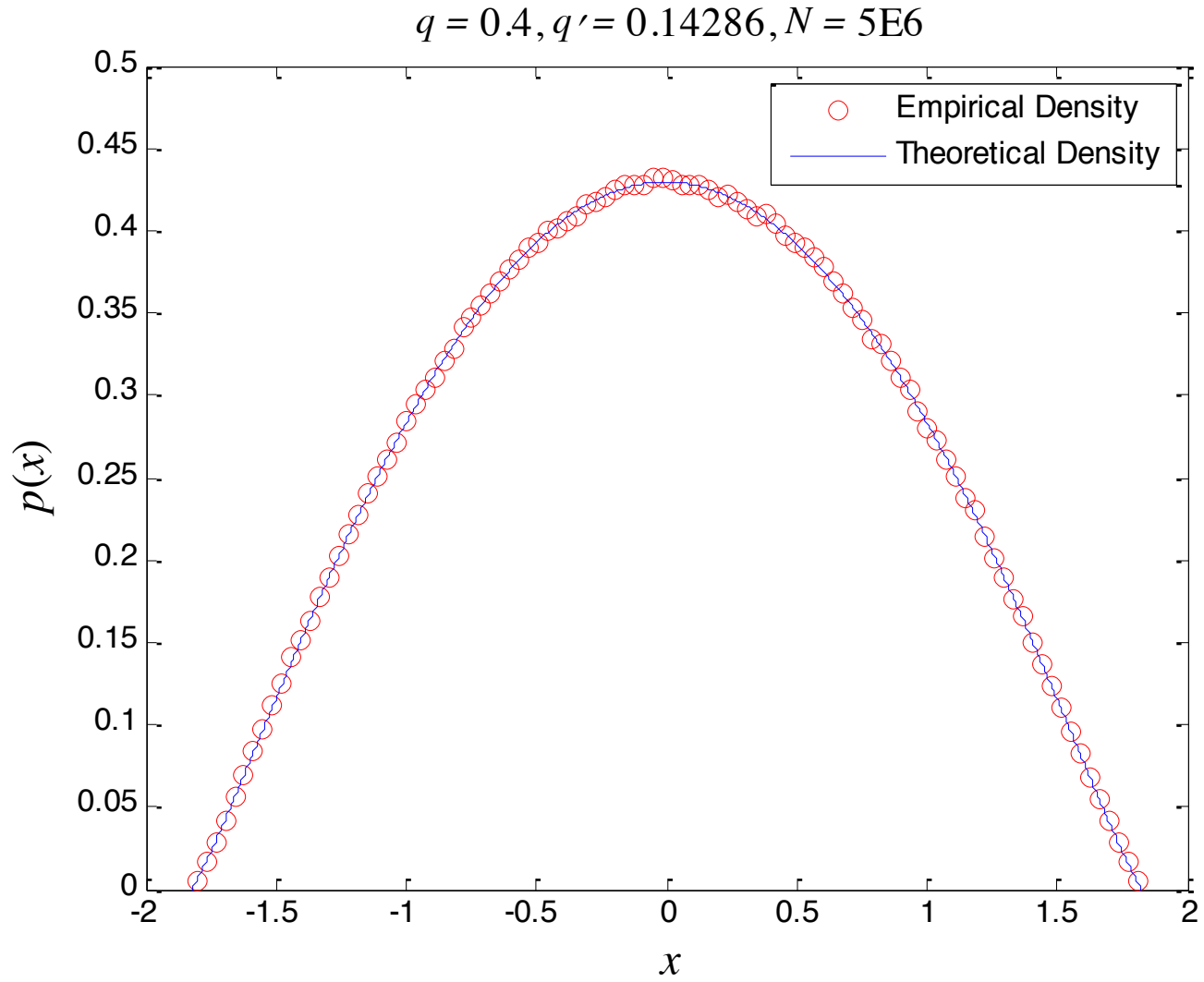

*Figure 2: Theoretical density* p(x) *and histogram of simulated data for a standard* q*-Gaussian distribution with* $0 < q' < 1$. *The distribution has compact support for this* $q'$.

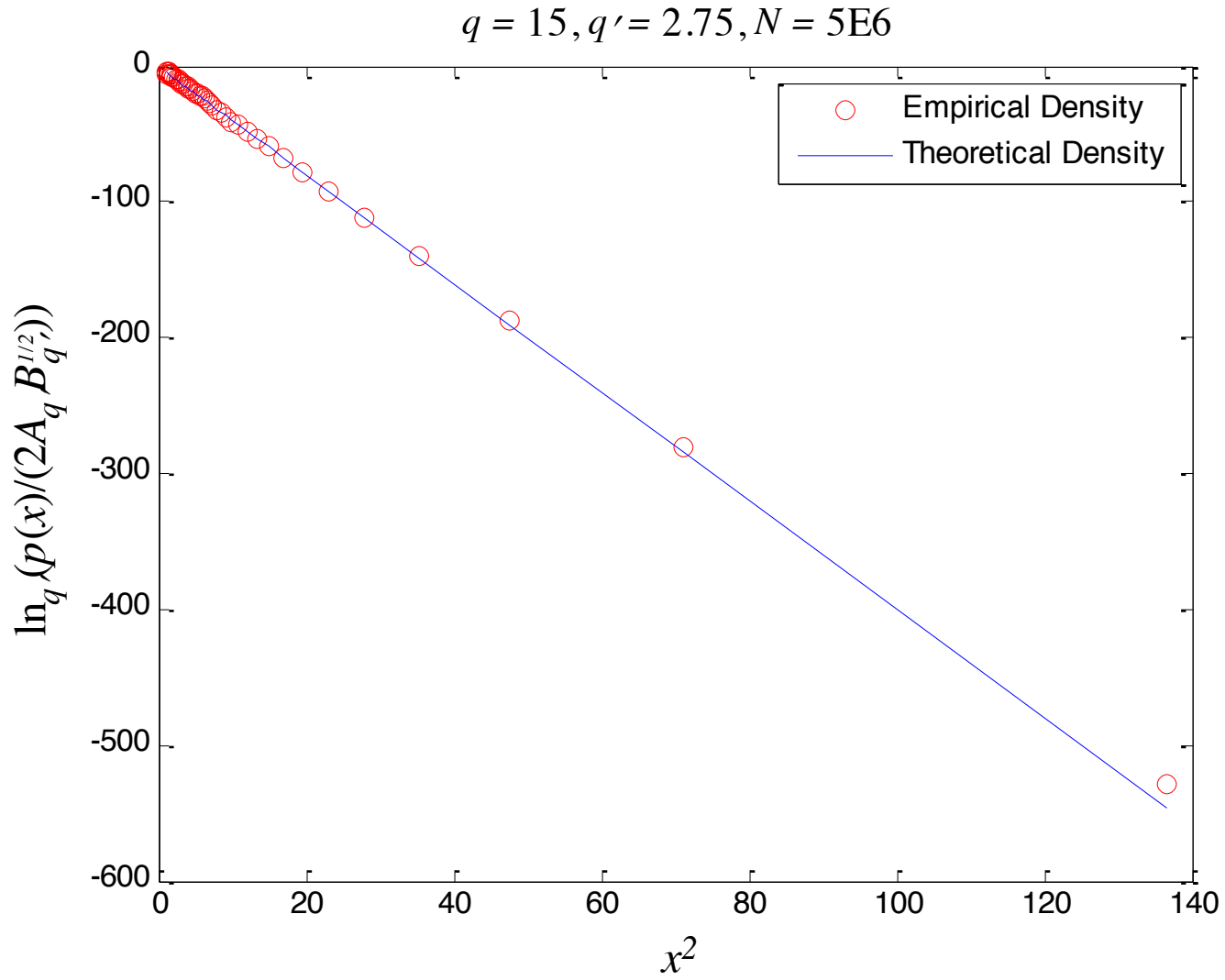

*Figure 4: Theoretical density* $\ln_{q'}\left(p(x)/A_{q'}\sqrt{B_{q'}}\right)$ *and histogram of simulated data for a standard* q*-Gaussian distribution, as a function of* $x^2$. *Plotted in this form, the* q*-Gaussian appears as a straight line. The value of* $q'$ *is the same as in Figure 3.*

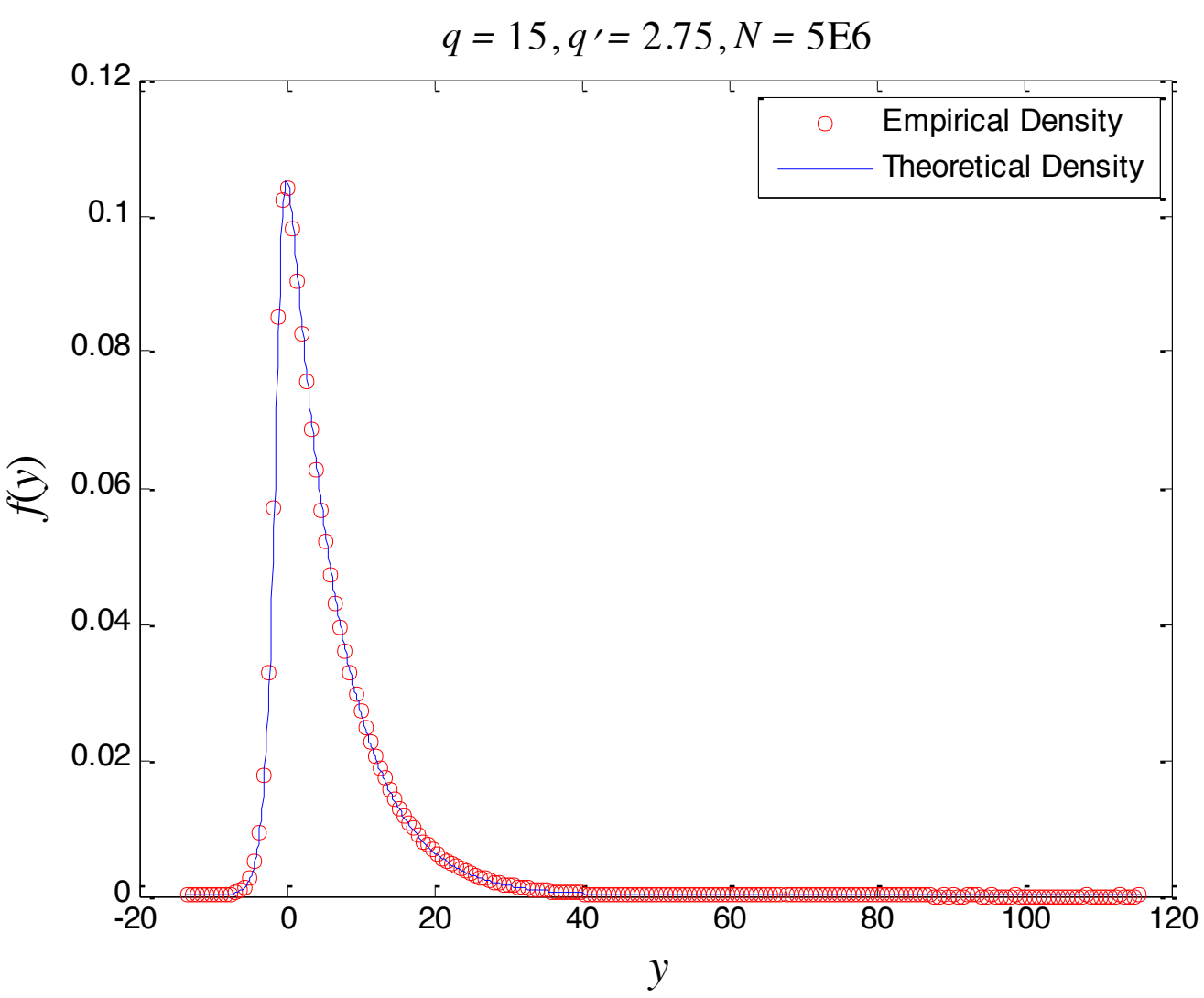

*Figure 5: Theoretical density* f(y) *of the random variable* $Y = \log(|X|)$*, and histogram of simulated data, for a standard* q-*Gaussian random variable* X*. The value of* $q'$ *is the same as in Figure 3.*

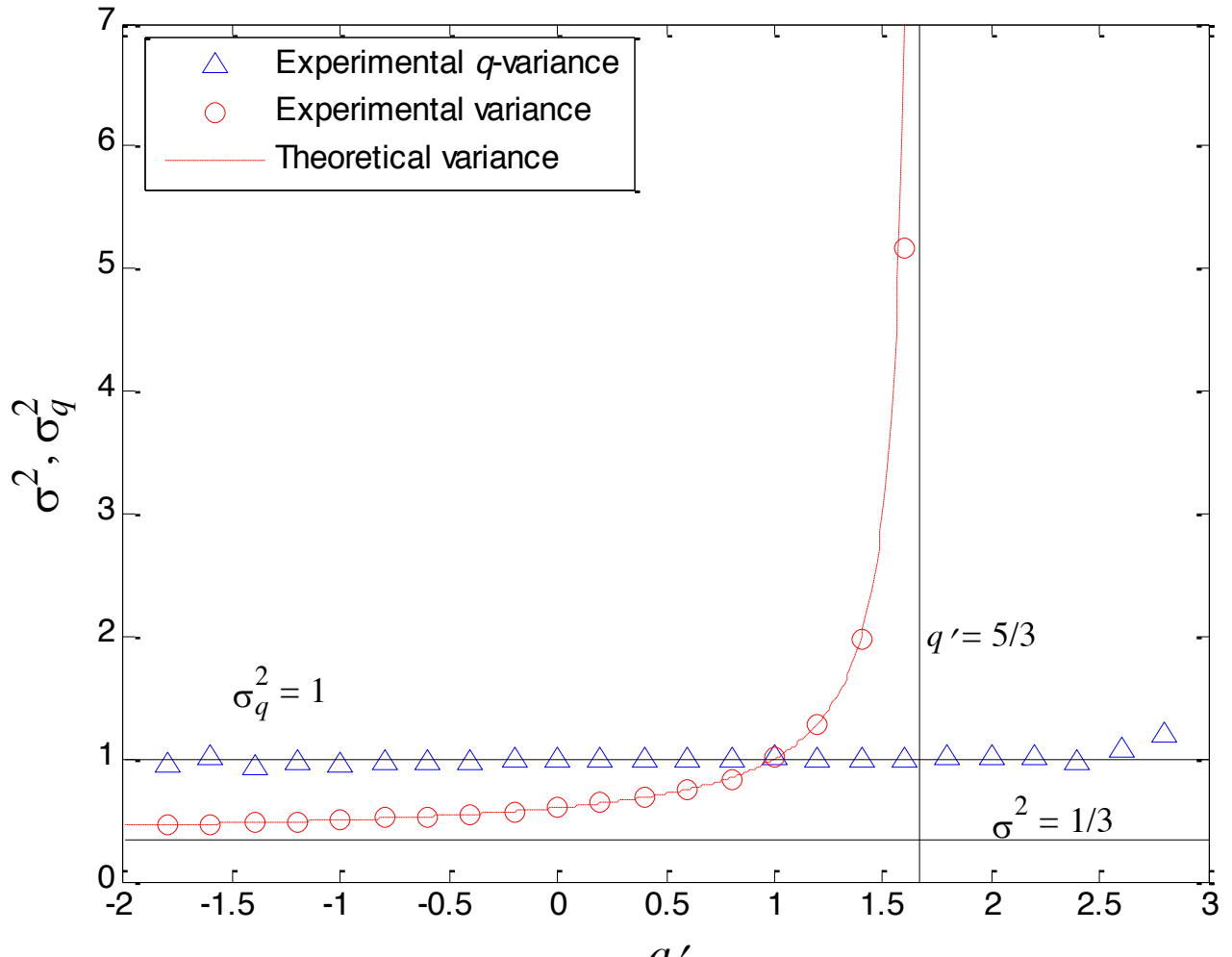

*Figure 6: Comparison of computed variance (circles) and* q-*variance (triangles). The usual second order central moment is seen to diverge for generated* $q' \geq 5/3$*. For* $q \to -\infty$ *the* q-*Gaussian approaches a uniform distribution on (-1,1) and so* $\sigma^2 \to 1/3$*. Very heavy tails for large* q *renders computation of* q-*variance difficult.*

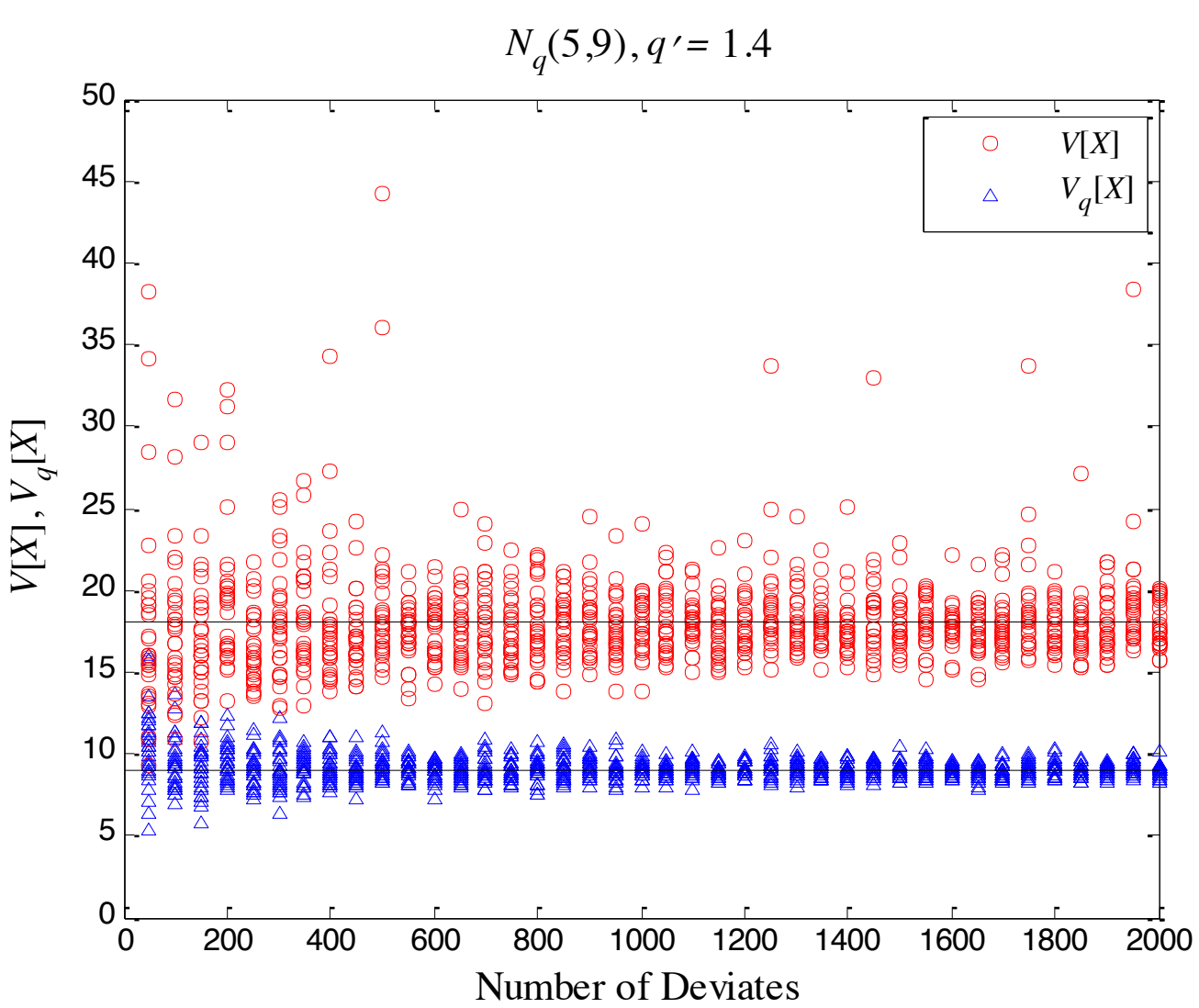

*Figure 7: Computed variance and* q-*variance for* q-*Gaussian distributions transformed by* $X = 5 + 3Z \sim N_q(5,9)$*. The* q-*variance is* $\bar{\sigma}_q^2 = 9$*, and since* $q' < 5/3$*, the variance is finite,* $\sigma^2 = 18$ *for this value of* $q'$*. Shown are the results from 30 runs at each sample size, where sample sizes range from 50 to 2000. The* q-*variance shows much faster convergence than the variance.*

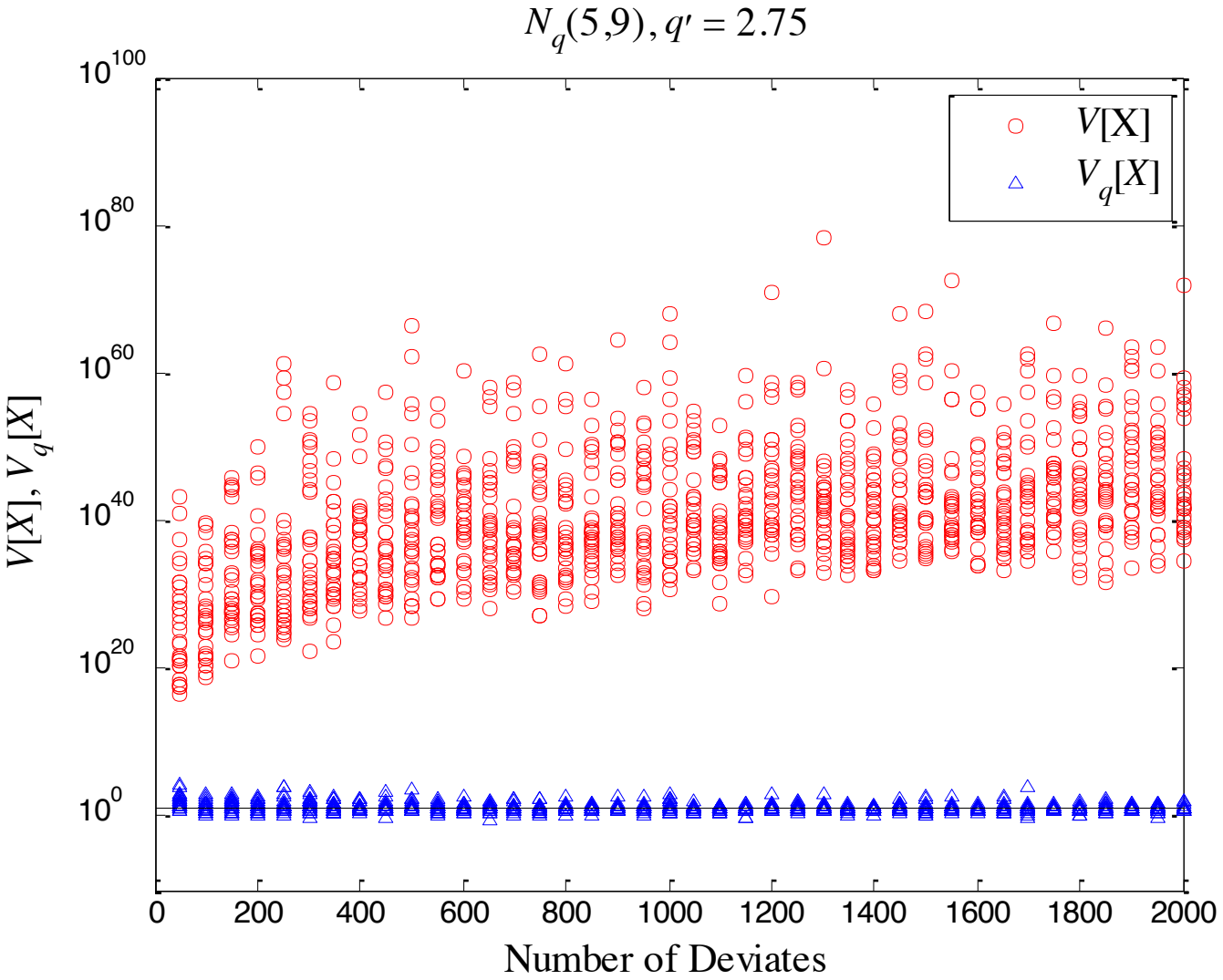

*Figure 8: Computed variance and* q-*variance for* q-*Gaussian distributions transformed by* $X = 5 + 3Z \sim N_q(5,9)$*. The* q-*variance remains finite, at the predicted value* $\bar{\sigma}_q^2 = 9$*. However, since* $q' > 5/3$*, the variance is infinite. Shown are the results from 30 runs at each sample size, where sample sizes range from 50 to 2000.*